\documentclass{emulateapj} 

\def\msun{M$_\odot$}
\def\mic{$\mu$m}
\def\den{$N_{\rm e}$}
\def\tem{$T_{\rm e}$}
\def\cm{cm$^{-3}$}

\shorttitle{IRS observations of the LMC planetary nebula SMP\,83}
\shortauthors{Bernard-Salas et al.}

\begin{document}

\title{IRS observations of LMC planetary nebula SMP\,83}




\author{J. Bernard-Salas\altaffilmark{1}}
\author{J.R. Houck\altaffilmark{1}}
\author{P.W. Morris\altaffilmark{2}}
\author{G.C. Sloan\altaffilmark{1}}
\author{S.R. Pottasch\altaffilmark{3}}
\author{D.J. Barry\altaffilmark{1}}

\email{jbs@isc.astro.cornell.edu}

\altaffiltext{1}{Center for Radiophysics and Space Research, Cornell
University, 219 Space Sciences Building, Ithaca, NY 14853-6801, USA}
\altaffiltext{2}{Spitzer Science Center, IPAC/California Inst. of
Technology, 1200 E. California Blvd., Pasadena, CA 91125, USA, \& NASA
Herschel Science Center, IPAC/Caltech, MS 100-22, Pasadena, CA 91125}
\altaffiltext{3}{Kapteyn Astronomical Institute, P.O. Box 800, 9700 AV
Groningen, The Netherlands}

\begin{abstract}
  
  The first observations of the infrared spectrum of the LMC planetary
  nebula SMP\,83 as observed by the recently launched Spitzer Space
  Telescope are presented.  The high resolution ($R\sim600$) spectrum
  shows strong emission lines but no significant continuum. The infrared
  fine structure lines are used, together with published optical
  spectra, to derive the electron temperature of the ionized gas for
  several ions. A correlation between the electron temperature with
  ionization potential is found. Ionic abundances for the observed
  infrared ions have been derived and the total neon and sulfur
  abundances have been determined.  These abundances are compared to
  average LMC abundances of \ion{H}{2} regions to better understand
  the chemical evolution of these elements. The nature of the progenitor
  star is also discussed.

\end{abstract}

\keywords{ISM: abundances --- ISM: lines and bands --- planetary
  nebulae: individual(SMP\,83) --- stars: evolution --- stars: Wolf--Rayet}

\section{Introduction}

Stars of low and intermediate mass evolve through the planetary nebula
phase. By the ejection of their outer layers the Planetary Nebulae
(PNe) contribute to the enrichment of the interstellar medium.
Spectroscopic studies of PNe are essential to better understand and
quantify this enrichment.  Until the launch of the Spitzer Space
Telescope it was impossible to fully study PNe in the infrared outside
the galaxy's chemical environment because of the faintness of
extragalactic targets.  Reaching targets in the Large and Small
Magellanic Clouds (LMC, SMC) is particularly appealing because they
have known distances and arise in a lower metallicity environment than
galactic targets. Since the distance is known the enrichment of
expelled nebular material can be combined with accurate knowledge of
the central star's luminosity.

In this letter, the infrared spectrum of the LMC planetary nebula
SMP\,83, taken with the IRS\footnote{The IRS was a collaborative
  venture between Cornell University and Ball Aerospace Corporation
  funded by NASA through the Jet Propulsion Laboratory and the Ames
  Research Center.} spectrograph on board the Spitzer Space Telescope,
is presented. The nature of the central star of SMP\,83 is not yet
well known and has been the subject of several studies.
\citet{dopita}, using HST imagery, conclude that SMP\,83 is a type~I
PN with a very massive progenitor star of 6~\msun, close to the
progenitor mass limit for a PN. \citet{torres} reported the sudden
development of Wolf-Rayet (WR) features in the \ion{He}{2} line at
$\lambda$4686 in the nebula. During 1993 and 1994 the central star
experienced an increase in luminosity which was studied by
\citet{vassiliadis} and more recently by \citet{hamann}. The latter
showed that the chemical composition of the atmosphere is that of
incompletely processed CNO material. They suggest several scenarios
for the nature of the central star, including a low-- or high--mass
single star or a low-- or high--mass binary system. \citet{hamann}
conclude that binary systems present the fewest contradictions when
explaining all the central star peculiarities. \citet{pena3} indicated
that the fast variations of the stellar parameters are similar to
those expected in the initial phase of the {\em born--again} scenario
in which stars undergo a final helium flash.

\citet{mea} presented optical spectroscopy for 30~PNe in the LMC
including SMP\,83.  They found that the densities obtained with the
\ion{S}{2} lines were slightly lower than those derived with the
\ion{O}{2} lines.  \citet{tsamis} made spectroscopic observations of
several galactic and LMC PNe, including SMP\,83.  Using collisionally
excited lines they derived densities, temperatures and ionic
abundances for the objects in their sample. \citet{pena2} and
\citet{pena3} combined optical and ultraviolet (IUE) data with
ionization models to derive the composition of the nebula.  They found
that the C and O abundances were under--abundant when compared to
H\,II regions in the LMC. \citet{pena3} based on the lack of evidences
of freshly synthesized carbon conclude that the third dredge--up did
not occur.

For the first time, complete mid--infrared spectra (from 5.3 to
40~\mic) for this nebula are available. The infrared region of the
spectrum is very useful for several reasons. The lines seen in the
infrared are not sensitive to the temperature, making them reliable
indicators of the chemical composition of the nebulae. The peak of
dust emission occurs in the infrared and features such as PAHs and
silicates (when present) can only be studied in this part of the
spectrum.


\section{Observations and data reduction}
\label{obs_s}

The observations were made with the IRS spectrograph \citep{houck} on
board the Spitzer Space Telescope \citep{werner}. The data were
obtained during the in-orbit checkout phase of the satellite, on
2003~Nov~9 (observation number 7857664). The diameter of the nebula is
$<2\arcsec$ and therefore can be calibrated as a point source.

The observations at SMP\,83 consisted of spectra from Short--High
(SH), Long-High (LH), and the first order of Short--Low (SL1) modules.
The wavelength coverage for each module is; SL1 (from 7.5--14.2~\mic),
SH (from 10--19.5~\mic), LH (from 19.3--37~\mic). The high--resolution
modules provide a spectral resolution of $\sim$600 and the
low--resolution $\sim$90.  The ``staring'' observation mode was used,
with two nod positions for each module. The observation times were
180, 240 and 480 seconds for SL1, SH and LH respectively.  These total
integration times represent the co--addition of shorter individual
ramps.

\begin{figure}
  \includegraphics[angle=0,width=8.5cm]{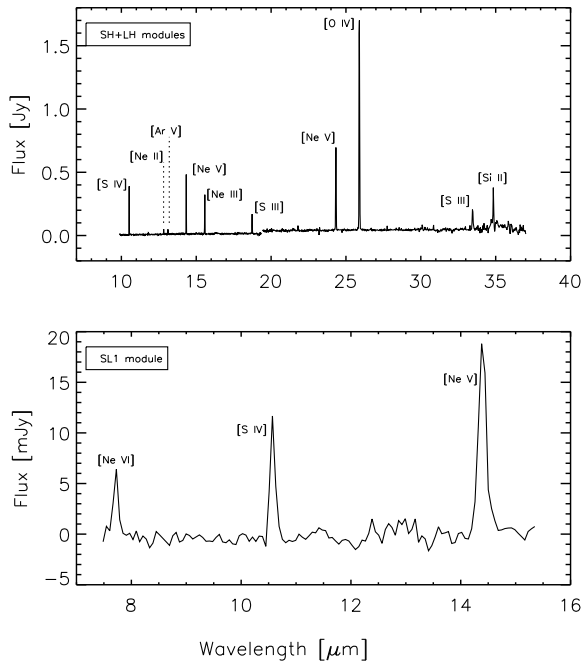}
  \caption{High (top) and low (bottom) resolution spectra of LMC
    planetary nebula SMP\,83. Note that the low resolution observation
    consisted only on the first order of the SL module
    (7.5--14.2~\mic).
\label{spec_f}}
\end{figure}

The data were processed through a copy of the Spitzer Science Center's
pipeline reduction software (version S9.1) maintained at Cornell.
From that point the reduction and extraction techniques were carried
out as follows: The mean of the flux estimates from each ramp cycle
were combined.  The two nod positions in SL1 were subtracted to
eliminate the contribution from the background.  The resulting images
were extracted using the Cornell-developed software package SMART
\citep{sarah}.  
The high--resolution spectra were
calibrated by dividing the extracted spectrum of the source by the
spectrum of the standard star HR\,6348 (observation number 7834624)
and multiplying by its template \citep{cohen}.  Remaining spikes were
removed manually.  The resulting spectra in SH and LH at the second
nod position are shown in Figure~\ref{spec_f}.  The quality of the
spectra are remarkable; it is the first detection of these lines in an
extragalactic PN.  Due to a mispointing the SH observation at the
first nod position was not used in the analysis of the data.

\section{Analysis of the emission lines}

The measured lines are listed in the third column of
Table~\ref{fluxabun_t}. Several calibration and cross-check techniques
were applied to the data. A comparison of the measured fluxes using
these different techniques leads us to conclude that the uncertainty
in the absolute flux is 20--25\%. The only exceptions are
\ion{[Ne}{6]} (this line is measured at the edge of the spectrum in
SL1 and is noisy) and the \ion{[Si}{2]} lines with uncertainties of
50\% and 30\% respectively.  Absolute flux calibration of the IRS
instrument is ongoing and these uncertainties will decrease with
improved future reductions.  The fine structure lines provide
information of the ionic abundances in the nebula, after the electron
density and temperature are determined.  Although the extinction
correction in the infrared is small the lines have been corrected for
extinction using $C(H\beta)=0.15$ \citep{mea,dopita} and are given in
column~4 of Table~\ref{fluxabun_t}.

  \subsection{Electron density and temperature}
   
  Intensity ratios of lines originating from levels close in energy
  are needed to derive the electron density \den.  Two lines of the
  same ion have been observed for \ion{Ne}{5} and \ion{S}{3}. However,
  these ions are sensitive to densities from $\sim$3000 to
  $\sim$10$^5$~\cm~and the observed ratios for both ions indicate that
  the density for this nebula is lower than 3000~\cm.  \citet{mea}
  using \ion{O}{2} and \ion{S}{2} ratios (which are sensitive to lower
  density regimes) derived densities of 2470 and 2220
  \cm~respectively. \citet{tsamis} obtain electron densities of 1900
  \cm~and 5700 \cm~for respectively \ion{S}{2} and \ion{Ar}{4}.  We
  have adopted for our calculations an average of those given by
  \citet{mea} (\den=2350 \cm). This density is low enough than the critical
  density of all the lines listed in Table\,\ref{fluxabun_t} (except
  for the 33.4 and 34.8 \mic~lines) and therefore does not
  affect the line intensities. If an average density of all the
  values given above is
adopted (3000 \cm) none of the conclusions would be affected.
  
  \begin{deluxetable}{c c c c c}[t]
    \tablewidth{8cm} \tablecaption{Measured fluxes in the high
      resolution spectra and calculated ionic abundances (relative to
      hydrogen).\label{fluxabun_t}} \tablehead{\colhead{$\lambda$} &
      \colhead{Ident.} & \multicolumn{2}{c}{Flux\tablenotemark{a}} &
      \colhead{$N_{\rm ion}$/$N_{\rm p}$} \\ 
      \cline{3-4} \colhead{\makebox{\rule[0cm]{0cm}{0.33 cm}[\mic]}} &
      \colhead{} & \colhead{Observed} & \colhead{De-reddened} &
      \colhead{( $\times$10$^{-5}$)}}

    \startdata
  
    7.652\tablenotemark{b} & \ion{[Ne}{6]}  &  8.53  & 8.56 &  1.09  \\
    10.52                  & \ion{[S}{4]}   &  14.7  & 15.9 &  0.19  \\ 
    12.82                  & \ion{[Ne}{2]}  &  1.55  & 1.56 &  1.15  \\ 
    13.11                  & \ion{[Ar}{5]}  &  1.23  & 1.24 &  0.02  \\
    14.33                  & \ion{[Ne}{5]}  &  14.2  & 14.3 &  1.14  \\
    15.55                  & \ion{[Ne}{3]}  &  10.3  & 10.4 &  3.72  \\  
    18.73                  & \ion{[S}{3]}   &  3.92  & 3.94 &  0.22  \\ 
    24.33                  & \ion{[Ne}{5]}  &  15.0  & 15.1 &  1.26  \\
    25.89                  & \ion{[O}{4]}   &  43.7  & 43.9 &  6.08  \\
    33.46                  & \ion{[S}{3]}   &  3.47  & 3.48 &  0.36  \\
    34.84                  & \ion{[Si}{2]}  &  2.76  & 2.77 &  0.56  \\
    
    \enddata
    
    \tablenotetext{a}{In units of 10$^{-14}$~erg~cm$^{-2}$s$^{-1}$.}
    \tablenotetext{b}{Measured from the low resolution spectrum and
      scaled so that the \ion{S}{4} and \ion{Ne}{5} lines measured in
      SL1 (module which was reduced and calibrated with SMART software)
      matched those fluxes in SH.}
  \end{deluxetable}

\begin{figure}
    \includegraphics[width=9cm]{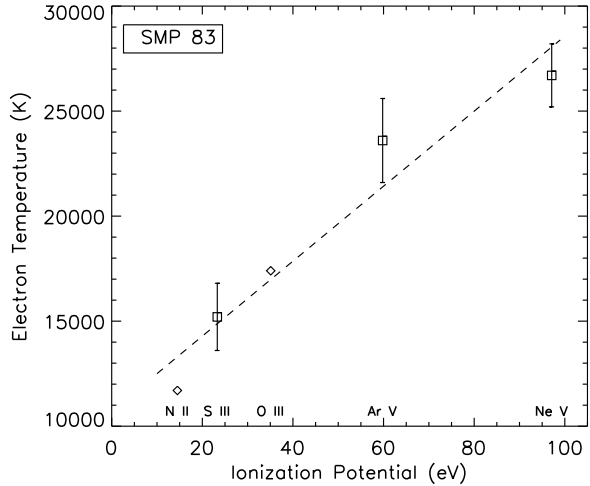}
    \caption{Derived electron temperature  versus the
      ionization potential require to reach the respective ionization
      level. The diamonds are the \tem~derived by \citet[][see
      text]{mea} and the squares those derived in this
      letter.\label{tem_f}}
  \end{figure}

  To derive the electron temperature \tem~ratios of lines originating
  from energy levels that differ by several electron volts are needed.
  We combine line strengths from our spectra with optical lines given
  by \citet{mea} to increase the numbers of ions available to derive
  \tem.  \citet{mea} derived \tem=17\,400 K for \ion{O}{3} and
  \tem=11\,700 K for \ion{N}{2} using the ratio of optical lines.
  Their optical measurements together with the infrared lines allow us
  to derive the temperature for three more ions, \ion{S}{3},
  \ion{Ar}{3}, and \ion{Ne}{5}. The ratios used are;
  18.7\,\mic/6312\,\AA~and 33.4\,\mic/6312\,\AA~for \ion{S}{3}, 13.1\,\mic/7006\,\AA~for
  \ion{Ar}{5}, and 14.3\,\mic/3426\,\AA~and 24.3\,\mic/3426\,\AA~for
  \ion{Ne}{5}. These temperatures together with those by \citet{mea}
  are presented in Figure~\ref{tem_f} versus the Ionization Potential
  (IP). The \ion{Ne}{5} temperature shown in Figure~\ref{tem_f} is an
  average of that found with the 14.3\,\mic/3426\,\AA~and
  24.3\,\mic/3426\,\AA~ratio.  The electron temperature clearly
  increases with IP and reaches very high values.  The dashed line in
  the figure represents the least--squares fit to the data.  High
  stages of ionization are probably formed close to the central star
  where the density of harder UV photons is higher, whereas low stages
  of ionization are probably located further away from the central
  star, thus giving lower temperatures.  This gradient has been
  observed in several galactic PNe \citep[e.g.][]{yo,pottasch}, but it
  is the first time that it is shown for a PN in the LMC. The
  abundances of the different ions have been calculated using a
  \tem~according to the IP of each ion given by the fit (dashed line)
  in Figure~\ref{tem_f}.
    
  \subsection{Ionic abundances}
  
  The ionic abundances have been derived using the de-reddened fluxes
  and \den~and \tem~discussed in the previous subsection. These are
  presented in the last column of Table~\ref{fluxabun_t}. In order to
  derive the ionic abundances with respect to hydrogen, we have made
  use of the observed H$\beta$ flux given by \citet{pena}
  ($\log(F_{\rm H\beta})=-12.99$). It is important to notice that
  \citet{webster} and \citet{mea2} give a much different observed
  value ($\log(F_{\rm H\beta})=-12.66$). It is unclear why these
  authors quote such different values.  \citet{mea2} and
  \citet{webster} derived the H$\beta$ flux using a filter centered on
  the line. It may be that they are also measuring part of the stellar
  flux. \citet{pena} measured the nebular H$\beta$ flux in three
  different epochs and obtain very similar values (the differences are
  only due to the smaller slit apertures used). The larger aperture
  used by \citet{pena} is big enough to contain the nebula.  Because
  of the consistency between the three values they quote and the fact
  that they separate the contribution from the star and the nebula,
  we have opted to use that of \citet{pena}. The derived \tem~in the
  previous section involves the combination of infrared and optical
  lines (where the latter are given relative to H$\beta$=100) and the
  good agreement seen for all the values supports the choice of the
  H$\beta$ flux made in this letter.  We note however that using the
  H$\beta$ flux from \citet{webster} or \citet{mea2} will bring the
  ionic abundances derived from the infrared lines down by a factor
  $\sim$2.
  
  SMP\,83 is a high excitation nebula; several lines have been
  measured allowing the determination of the ionic abundance of
  several elements.  The \ion{O}{4} abundance is high and is similar
  to that found by \citet{pena3}. It accounts for a third part of
  the total oxygen abundance measured by \citet{pena3}.  Almost all
  important stages of ionization in neon (except \ion{Ne}{4}) and
  sulfur (except \ion{S}{2}) have been measured.  To account for the
  unobserved stages of ionization the optical lines at $\lambda$4729
  and $\lambda$4728 (for \ion{Ne}{4}) by \citet{tsamis} and at
  $\lambda$6717 and $\lambda$6731 (for \ion{S}{2}) by \citet{mea} have
  been used. By adding the abundances of different stages of
  ionization, the total abundances of neon and sulfur are found to be
  8.6$\times10^{-5}$ and 4.8$\times10^{-6}$ respectively. Most of the
  absolute error in these abundances comes from the uncertainty in the
  fluxes \citep{yo2} and therefore the uncertainties in the abundances
  are $\sim$35\,\%.
  
  It is interesting to compare the neon and sulfur abundances to
  average LMC abundances of H\,II regions, [S/H]=5.2$\times10^{-6}$
  and [Ne/H]=5.4$\times10^{-5}$ \citep{ronald}. \citet{pena3} noticed
  with this comparison that the sulfur and neon abundances they
  derived where systematically lower than those of the LMC \ion{H}{2}
  regions they compared to.  In this paper the sulfur abundance in
  SMP\,83 agrees with that of the LMC.  This is important because
  \citet{marigo} pointed out that the sulfur abundance of the galactic
  PNe that they studied was lower than solar.  They suggested that
  either the solar sulfur abundance was overestimated or that sulfur
  could be destroyed in the course of evolution, contrary to
  evolutionary modeling.  Since the sulfur abundance derived in this
  paper is consistent with that of the LMC, there is support for the
  hypothesis of an overestimated solar sulfur abundance.  It would be
  interesting to derive the sulfur abundance for more LMC PNe to
  confirm this.  The neon abundance found here is higher than in
  previous studies \citep{pena3,dopita,tsamis}. A small part of the
  difference comes from the ionic abundances of \ion{Ne}{2} and
  \ion{Ne}{6} whose contributions were neglected in previous studies
  since it was not observed. Nonetheless, the infrared data show that
  its contribution to the total neon abundance is important. The
  larger difference however comes from the contribution from
  \ion{Ne}{3}.  In this letter this contribution is found to be 2
  times higher than that derived by \citet{tsamis} and \citet{pena3}
  who used the optical lines at 3868 and 3967~\AA.  This difference is
  similar to what has been seen in other nebulae
  \citep[i.e.][]{pottasch,yo2}.  The problem depends on the adopted
  \tem. The \tem~obtained when using the ratio of the 15.55\,\mic~and
  3869~\AA~lines is 12\,000 K. There are several galactic nebulae for
  which these lines give a much lower temperature than that of the
  \ion{[O}{3]} lines (i.e., NGC\,6302, NGC\,6537, NGC\,7027). If a
  higher temperature is used the ionic abundances derived from the
  optical lines decrease. However the low--level infrared lines are
  not affected by this problem since they hardly depend on the
  temperature and give better estimates of the ionic abundance.  The
  neon abundance seems larger than that in the LMC, although this is
  arguable considering the errors.  If real, this enrichment must have
  taken place likely via $\alpha$-captures starting from $^{14}$N
  during dredge-up. This mechanism marginalizes the
  $^{22}{\rm{Ne}}(\alpha,n)^{25}{\rm{Mg}}$ reactions expected to occur
  during thermal pulses, and extinguishes the He-burning shell.  This
  could explain the lack of carbon at the surface of SMP83
  \citep{hamann}.

  \subsection{Nature of the central star}
  
  A confirmation of the neon enrichment could lead to important constraints in
  the nature of the progenitor star of SMP\,83.  In massive stars
  ($M{\rm _{init}} > 25 M_\odot$), which are convectively unstable,
  the Ne production occurs through the chain reaction
  $^{18}{\rm{O}}(\alpha,n)^{20}{\rm{Ne}}$ and the observed abundances
  will depend on whether further $\alpha$ capture through
  $^{22}{\rm{Ne}}(\alpha,n)^{26}{\rm{Mg}}$ has occurred, and on core
  enrichment from the surface by rotational mixing \citep{meynet}.
  Neon enhancement in the massive stars thus coincides with carbon and
  oxygen enhancements and hydrogen depletion in the core. This leads
  spectroscopically at the surface to WC-type Wolf-Rayet stars,
  according to the evolutionary models and observational tests of
  massive stars in early and later post-main sequence stages
  \citep{morris1,morris2,dessart}.  Thus, if the neon enrichment is
   confirmed, the abundance pattern in the nebula and at the surface of
  SMP83 most likely excludes the central star as a Population I
  Wolf-Rayet star, as one of the scenarios proposed by \citet{hamann},
  and is more consistent with an alternate hypothesis of an
  intermediate mass star that has experienced efficient dredge-up in
  the post-AGB phase.

\section{Summary and conclusions}

Mid infrared spectra of LMC planetary nebula SMP\,83 have been
presented for the first time using the IRS spectrograph on board the
Spitzer Space Telescope.  The spectrum of SMP\,83 shows high
excitation lines and no signs of dust or a continuum associated with
the nebula.

The emission lines have been used with optical spectra from the
literature to derive the electron temperature for several ions in the
nebula. There is a correlation of the electron temperature with
ionization potential. The abundances for several ions have been
derived.  The total neon abundance, with additional contributions from
states seen only in the infrared, is higher than the
optically-determined abundance, which is more sensitive to the
temperature.  The neon abundance seems larger 
 than the adopted average neon LMC abundance. This could
suggests that some neon enrichment must have taken place in the course
of evolution. If the neon enrichment is confirmed, this could
favor the scenario of SMP\,83 having evolved from an intermediate mass
star instead of a massive Wolf--Rayet star as some studies have
suggested.  The sulfur abundance  agrees with that
of the LMC.  This might indicate that the problem of the low sulfur
abundance found in galactic PNe relative to solar comes from an
overestimation of the latter, but this should be confirmed with more
studies on the sulfur abundance in PNe in the LMC.

\acknowledgments We wish to thank the referee and M.Pe\~na for
clarifying us that the H$\beta$ flux given in \citet{pena} is not
corrected for extinction.  This work is based on observations made
with the Spitzer Space Telescope, which is operated by the Jet
Propulsion Laboratory, California Institute of Technology under NASA
contract 1407. Support for this work was provided by NASA through
Contract Number 1257184 issued by JPL/Caltech.


\end{document}